\documentclass[aps,twocolumn,pra,showpacs,floatfix]{revtex4}
\usepackage{epsfig}
\usepackage{graphicx}
\usepackage{dcolumn}
\usepackage{amsmath}

\begin{document}


\title{Relativistic corrections to transition frequencies of
Ag~I, Dy~I, Ho~I, Yb~II, Yb~III, Au~I and Hg~II
and search for variation of the fine structure constant}

\author{V. A. Dzuba and V. V. Flambaum}
\affiliation{School of Physics, University of New South Wales,
Sydney 2052, Australia}

\date{\today}

\begin{abstract}
Dependence of transition frequencies on the fine structure constant $\alpha=e^2/\hbar c$ is calculated
 for several many-electron
systems which are used or planned to be used in the laboratory search
for the time variation of the fine structure constant. In systems with a
large number of  electrons in open shells (from 11 to 15)  the relative effects of the variation
may be strongly enhanced. For the transitions which were
considered before the results are in good agreement with previous
calculations.

\end{abstract}

\pacs{PACS: 31.30.Jv, 06.20.Kr, 95.30.Dr}

\maketitle

\section{Introduction}

Theories unifying gravity with other interactions as well as many
cosmological models allow for space-time variation of fundamental
constants. Experimental search for the manifestation of this
variation spans the whole lifetime of the Universe from
Big Bang nucleosynthesis to the present-day very precise
atomic clock experiments (see, e.g. reviews~\cite{Uzan,Flambaum07a}).
An evidence that the fine-structure constant might be smaller
about ten billion years ago was found in the analysis of quasar absorption
spectra~\cite{Webb99,Webb01,Murphy01a,Murphy01b,Murphy01c,Murphy01d}.
This finding together with progress in developing of very precise
atomic frequency standards motivated many laboratory searches for
the present-day time variation of the fundamental constants
(see, e.g.~\cite{Lea}).
In particular, strong limit on the rate of the time variation of
the fine structure constant $\alpha$ ($\alpha=e^2/\hbar c$)
were found by comparing frequencies of different atomic
transitions over few years~\cite{Lea}.

Apart from the microwave atomic clocks and optical frequency standards,
a number of atomic transitions in which the change of frequency  due to
change of $\alpha$ is strongly enhanced has been suggested in
Refs.~\cite{Dzuba99a,Dzuba05,Angstmann06}.

Interpretation and planning of the measurements of the  $\alpha$ variation
require  atomic calculations to relate the change of  atomic frequencies  to the change
of the fine structure constant. A number of such calculations for
atomic optical transitions have been performed in our early
works~\cite{Dzuba99a,Dzuba05,Angstmann06,Dzuba00,Dzuba03,Angstmann04}.
Independent calculations for some optical transitions have been
recently reported in Ref.~\cite{Borsch}.

From the computational point of view the most important parameter
of an atom which determines the choice of the computational method
as well as the accuracy which can be achieved in the calculations is the
number of electrons in open shells. The larger the number the
more difficult are the calculations. Many optical frequency standards
are based on atoms or ions with just one or two valence electrons~\cite{Lea}.
Calculations for such systems are accurate and
reliable~\cite{Dzuba99a,Dzuba00,Angstmann04}.
However, many atomic systems which are used or planned to be used
in laboratory search for variation of the fine structure constant
have more then ten electrons in open shells. For example, strong limits
on the variation of alpha in time\cite{Budker07} and variation
of alpha due to change of the gravitation potential\cite{Budker07a}
were obtained with the use of dysprosium atom which has twelve
external electrons  
(see also~\cite{Dzuba99a,Dzuba03,Budker04}). There are plans to use
holmium (13 electrons) for similar measurements~\cite{Saffman}.
There are ongoing measurements or plans for measurements for
Ag~I~\cite{AgI}, Yb~II,~\cite{YbIIa,YbIIb}, Yb~III~\cite{YbIII} and Hg~II~\cite{HgII}
(see also a review~\cite{Lea} and references therein).
These systems involve states with excitations from $d$ or $f$ subshells
and therefore must be treated as many-valence-electrons systems.

Calculations for many-valence-electron atoms are difficult due
to the fast growth of the matrix size of the configuration interaction (CI)
eigenvalue problem with the increase of the single-electron basis.
In our recent paper on Fe~I\cite{Dzuba07} we used a version of the CI
method which is similar to multi-configuration CI method (see,
e.g. Ref.~\cite{Grant})
and which allows to obtain reasonably accurate result with a very short basis.
In present paper we use this method for many-electron systems which are
of the interest for laboratory search of the variation of the fine structure
constant. The aim of the calculation is to check our early results as well
as to calculate relativistic energy shifts for atomic transitions which
have never been considered before.

\section{Method}

Detailed discussion of the method can be found in our early
works~\cite{Dzuba99a,Dzuba07}.
Here we repeat its major points.

It is convenient to present the dependence of atomic frequencies on
the fine-structure constant $\alpha$ in the vicinity of its physical
value $\alpha_0$ in the form
\begin{equation}
  \omega(x) = \omega_0 + qx,
\label{omega}
\end{equation}
where $\omega_0$ is the present laboratory value of the frequency and
$x = (\alpha/\alpha_0)^2-1$, $q$ is the coefficient which is to be
found from atomic calculations. Note that
\begin{equation}
 q = \left .\frac{d\omega}{dx}\right|_{x=0}.
\label{qq}
\end{equation}
To calculate this derivative numerically we use
\begin{equation}
  q \approx  \frac{\omega(+\delta) - \omega(-\delta)}{2\delta}.
\label{deriv}
\end{equation}
In the present calculations we use $\delta = 0.05$, which leads to
\begin{equation}
  q \approx  10 \left(\omega(+0.05) - \omega(-0.05)\right).
\label{deriv05}
\end{equation}
In a single-electron approximation relativistic energy shift
can be estimated using the formula~\cite{Dzuba99a}
\begin{equation}
  \Delta_a = \frac{E_a}{\nu_a}(Z\alpha)^2 \left[\frac{1}{j_a+1/2}-C(Z,j_a,l_a)\right],
\label{delta}
\end{equation}
where $a$ is the index for a single-electron state, $E_a$ is its energy,
$\nu_a$ is its effective principal quantum number ($\nu_a = 1/\sqrt{-2E_a}$),
$j_a$ and $l_a$ are total and angular momenta of the state $a$.
$C(Z,j_a,l_a)$ is a parameter which is introduced to simulate the
effect of Hartree-Fock exchange interaction and other many-body
effects.
For a transition between many-electron states which can be approximated
as a single-electron transition from state $a$ in lower level to state
$b$ in upper level one has
\begin{equation}
  q \approx \Delta_b - \Delta_a.
\label{qab}
\end{equation}
The formulas (\ref{delta}),(\ref{qab}) are too inaccurate for practical
use in the interpretation
of the measurements. However, they are very useful for predicting what one
can expect to find in different atomic transitions and for explaining
the values and sign of the relativistic corrections.
We will use it for the discussion of our results.

For accurate numerical calculations of the coefficients $q$ using
(\ref{deriv05}),  $\alpha$ must be varied
in the computer code.  Therefore, it is convenient to use a form
of the single electron wave function in which the dependence on $\alpha$ is
explicitly shown (we use atomic units in which $e=\hbar=1, \alpha = 1/c$)
\begin{equation}
    \psi(r)_{njlm}=\frac{1}{r}\left(\begin {array}{c}
    f_{v}(r)\Omega(\mathbf{n})_{\mathit{jlm}}  \\[0.2ex]
    i\alpha g_{v}(r)  \widetilde{ \Omega}(\mathbf{n})_{\mathit{jlm}}
    \end{array} \right),
\label{psi}
\end{equation}
where $n$ is the principal quantum number and an index $v$
replaces the three-number set $n,j,l$.
This leads to a form of radial equation for single-electron
orbitals which also explicitly depends on $\alpha$:
\begin{equation}
    \begin {array}{c} \dfrac{df_v}{dr}+\dfrac{\kappa_{v}}{r}f_v(r)-
    \left[2+\alpha^{2}(\epsilon_{v}-\hat{V}_{HF})\right]g_v(r)=0,  \\[0.5ex]
    \dfrac{dg_v}{dr}-\dfrac{\kappa_{v}}{r}f_v(r)+(\epsilon_{v}-
    \hat{V}_{HF})f_v(r)=0, \end{array}
\label{Dirac}
\end{equation}
here $\kappa=(-1)^{l+j+1/2}(j+1/2)$,
and $\hat{V}_{HF}$ is the Hartree-Fock potential.
Equation (\ref{Dirac}) with $\alpha = \alpha_0 \sqrt{\delta +1}$
and different Hartree-Fock potential $\hat{V}_{HF}$ for
different configurations is used to construct single-electron orbitals.

\begin{table}
\caption{Configurations and effective core polarizabilities ($\alpha_p$, a.u.)
used in the calculations.}
\label{sets}
\begin{ruledtabular}
  \begin{tabular}{l c c r l l l}
\multicolumn{1}{c}{Atom} & $Z$ & $N_v$\footnotemark[1] &
\multicolumn{1}{c}{Set} &
\multicolumn{1}{c}{Parity} &
\multicolumn{1}{c}{Configuration} &
\multicolumn{1}{c}{$\alpha_p$} \\
\hline
Ag~I & 47 & 11 &  1 & Even &  $4d^{10}5s$    & 0.4 \\
     &    &    &  2 & Even &  $4d^{9}5s^2$   & 0.414 \\
Dy~I & 66 & 12 &  1 & Even &  $4f^{10}6s^2$  & 0.4 \\
     &    &    &  2 & Even &  $4f^{10}5d6s$  & 0.397 \\
     &    &    &  3 & Even &  $4f^{9}6s^26p$ & 0.4039 \\
     &    &    &  4 & Even &  $4f^{9}5d6s6p$ & 0.389 \\
     &    &    &  5 & Odd  &  $4f^{9}5d^26s$ & 0.3895 \\
     &    &    &  6 & Odd  &  $4f^{9}5d6s^2$ & 0.4    \\
     &    &    &  7 & Odd  &  $4f^{10}6s6p$ & 0.393 \\
Ho~I & 67 & 13 &  1 & Odd  &  $4f^{11}6s^2$   & 0.4 \\
     &    &    &  2 & Odd  &  $4f^{10}6s^26p$ & 0.401 \\
     &    &    &  3 & Odd  &  $4f^{11}5d6s$   & 0.401 \\
     &    &    &  4 & Odd  &  $4f^{10}5d6s6p$ & 0.39 \\
     &    &    &  5 & Odd  &  $4f^{11}6p^2$   & 0.39 \\
     &    &    &  6 & Even &  $4f^{10}5d6s^2$ & 0.3927 \\
     &    &    &  7 & Even &  $4f^{11}6s6p$   & 0.3962 \\
     &    &    &  8 & Even &  $4f^{10}5d^26s$ & 0.39 \\
     &    &    &  9 & Even &  $4f^{10}5d6p^2$ & 0.4 \\
     &    &    & 10 & Even &  $4f^{10}6s6p^2$ & 0.4 \\
Yb~II & 70 & 15 & 1 & Even &  $4f^{14}6s$     & 0.4 \\
      &    &    & 2 & Even &  $4f^{13}6s^2$   & 0.399 \\
      &    &    & 3 & Even &  $4f^{13}5d6s$   & 0.3911 \\
      &    &    & 4 & Even &  $4f^{13}5d^2$   & 0.39 \\
Yb~III & 70 & 14 & 1 & Even &  $4f^{14}$     & 0.4 \\
       &    &    & 2 & Even &  $4f^{13}5d$   & 0.3914 \\
       &    &    & 3 & Even &  $4f^{13}6s$   & 0.3977 \\
Au~I   & 79 & 11 & 1 & Even &  $5d^{10}6s$    & 0.4 \\
       &    &    & 2 & Even &  $5d^{9}6s^2$   & 0.417 \\
Hg~II & 80 & 11 & 1 & Even &  $5d^{10}6s$    & 0.4 \\
      &    &    & 2 & Even &  $5d^{9}6s^2$   & 0.426 \\
\end{tabular}
\footnotetext[1]{$N_v$ is the number of valence electrons.}
\end{ruledtabular}
\end{table}

Table~\ref{sets} lists configurations considered in present work.
For Ag~I, Au~I and Hg~II we use only ground-state configuration
and configurations, involving excitation from the upper core $d$-state.
The latter corresponds to the states which are to be used in the measurements.
We add more configurations for Yb~II and Yb~III and even more for
Dy~I and Ho~I. In the latter atoms the states of interest are highly
excited ones for which configuration mixing is strong and should
 be taken into account more accurately.

The self-consistent Hartree-Fock procedure is done for every configuration
listed in Table~\ref{sets} separately. Then valence states found in the
Hartree-Fock calculations are used as basis states for the CI calculations.
It is important for the CI method that the atomic core remains
the same for all configurations. We use the core which corresponds to the
ground state configuration. Change in the core due to change of the valence
state is small and can be neglected. This is because core states are not
sensitive to the potential from the electrons which are on large distances
(like $6s$, $6p$ and $5d$ electrons). The $4f$ electrons are on smaller distances
and have larger effect on atomic core. However, in all the cases
(see Table~\ref{sets}) only one among about ten $4f$ electrons change its state.
Therefore their effect on atomic core is also small. More detailed
discussion of the effect of valence electrons on atomic core can be
found in Refs.~\cite{VN,VN1}.

The effective Hamiltonian for $N_v$ valence electrons has the form
\begin{equation}
  \hat H^{\rm eff} = \sum_{i=1}^{N_v} \hat h_{1i} +
  \sum_{i < j}^{N_v} e^2/r_{ij},
\label{heff}
\end{equation}
$\hat h_1(r_i)$ is the one-electron part of the Hamiltonian
\begin{equation}
  \hat h_1 = c \mathbf{\alpha \cdot p} + (\beta -1)mc^2 - \frac{Ze^2}{r}
 + V_{core} + \delta V.
\label{h1}
\end{equation}
Here $\mathbf{\alpha}$ and $\beta$ are Dirac matrixes, $V_{core}$ is
Hartree-Fock potential due to core electrons
and $\delta V$
is the term which simulates the effect of the correlations between core
and valence electrons. It is often called {\em polarization potential} and
has the form
\begin{equation}
  \delta V = - \frac{\alpha_p}{2(r^4+a^4)}.
\label{dV}
\end{equation}
Here $\alpha_p$ is polarization of the core and $a$ is a cut-off parameter
(we use $a = a_B$).

The form of the $\delta V$ is chosen to coincide with the standard polarization
potential on large distances ($-\alpha_p/2r^4$). However we use it on distances
where valence electrons are localized. This distances are not large, especially
for the $4f$ electrons. Therefore we consider $\delta V$ as only rough
approximation to real correlation interaction between core and valence
electrons and treat $\alpha_p$ as fitting parameters. The values of $\alpha_p$
for each configuration of interest are presented in Table~\ref{sets}.
They are chosen to fit the experimental position of the configurations
relative to each other. For all configurations of the same atom the values of
$\alpha_p$ are very close. This is not a surprise since the core is
nearly the same for all configurations of interest.
One can probably say that small difference in $\alpha_p$
for different configurations simulates the effect of incompleteness of the
basis and other imperfections in the calculations.

\begin{table}
\caption{Energy levels (cm$^{-1}$) and $g$-factors of some low
states of Dy~I}
\label{Dy-en}
\begin{ruledtabular}
  \begin{tabular}{l c r l r l}
Conf. & $J$ & \multicolumn{2}{c}{Experiment\footnotemark[1]} &
 \multicolumn{2}{c}{Calculations} \\
 & & \multicolumn{1}{c}{Energy} & \multicolumn{1}{c}{$g$} &
       \multicolumn{1}{c}{Energy} & \multicolumn{1}{c}{$g$} \\
\hline
$4f^{10}6s^2$  &  8 &     0.00~ & 1.242 &     0 & 1.2428 \\
               &  7 &  4134.23~ & 1.173 &  4409 & 1.1747 \\
               &  6 &  7050.61~ & 1.072 &  7600 & 1.0723 \\
               &  5 &  9211.58~ & 0.911 &  9983 & 0.9080 \\
               &  4 & 10925.25~ & 0.618 & 11840 & 0.6163 \\
$4f^{10}5d6s$  &  9 & 17514.50~ & 1.316 & 17703 & 1.3145 \\
               &  8 & 18903.21~ & 1.22  & 19556 & 1.2754 \\
               &  7 & 21074.20~ & 1.24  & 21881 & 1.1983 \\
               &  8 & 17613.36~ & 1.33  & 17871 & 1.3300 \\
               &  7 & 18937.78~ & 1.28  & 19633 & 1.3012 \\
               &  6 & 21159.79~ & 1.24  & 22042 & 1.2116 \\
               &  7 & 18094.52~ & 1.38  & 18308 & 1.3835 \\
               &  6 &          &       & 20090 & 1.3078 \\
               &  5 &          &       & 22478 & 1.2198 \\
               & 10 & 18462.65~ & 1.282 & 18461 & 1.2883 \\
               &  9 & 19240.82~ & 1.217 & 19592 & 1.2277 \\
               &  8 & 20193.60~ & 1.16  & 20893 & 1.1700 \\
$4f^{10}6s^2$  &  8 & 19019.15~ & 1.14  & 21377 & 1.1113 \\
$4f^{10}5d6s$  & 11 & 19348.72~ & 1.27  & 19295 & 1.2675 \\
               & 10 & 19797.96\footnotemark[2]
                               & 1.21  & 20077 & 1.2089 \\ 
               &  9 & 20209.00~ & 1.14  & 20847 & 1.1261 \\
$4f^{9}6s^26p$ &  7 & 20614.32~ & 1.32  & 19835 & 1.3372 \\
               &  8 & 20789.85~ & 1.32  & 19832 & 1.2997 \\
$4f^{10}5d6s$  &  8 & 21603.04~ & 1.26  & 23205 & 1.2514 \\
               &  7 & 21778.43~ & 1.26  & 23232 & 1.2419 \\
               &  9 & 22045.79~ & 1.22  & 23429 & 1.2677 \\
               & 10 & 22487.14~ & 1.197 & 24132 & 1.2162 \\
$4f^{9}5d6s6p$ &  8 & 23031.46~ & 1.37  & 23132 & 1.3730 \\
$4f^{9}5d6s^2$ & 10 & 12892.76~ & 1.29  & 12920 & 1.2933 \\
$4f^{10}6s6p$  & 10 & 17513.33~ & 1.30  & 17582 & 1.2944 \\
$4f^{9}5d^26s$ & 10 & 19797.96\footnotemark[2]
                               & 1.367 & 19693 & 1.3677 \\ 
$4f^{9}5d^26$  & 10 & 21788.93~ & 1.34  & 22312 & 1.3340 \\
\end{tabular}
\footnotetext[1]{NIST, Ref.~\cite{NIST}}
\footnotetext[2]{States used in the measurements~\cite{Budker07,Budker07a}}
\end{ruledtabular}
\end{table}

\begin{table}
\caption{Energy levels (cm$^{-1}$) and $g$-factors of some low
states of Ho~I}
\label{Ho-en}
\begin{ruledtabular}
  \begin{tabular}{l l c r r l}
Conf. & Parity & $J$ & \multicolumn{1}{c}{Expt.\footnotemark[1]} &
 \multicolumn{2}{c}{Calculations} \\
 & & & \multicolumn{1}{c}{Energy} &
       \multicolumn{1}{c}{Energy} & \multicolumn{1}{c}{$g$} \\
\hline
$4f^{11}6s^2$   & Odd  & 15/2 &     0.00 &     0 & 1.20 \\
                &      & 13/2 &  5419.70 &  5770 & 1.11 \\
$4f^{10}6s^26p$ & Odd  & 15/2 & 18572.28 & 18343 & 1.28 \\
$4f^{11}5d6s$   & Odd  & 13/2 & 18867.40 & 18684 & 1.37 \\
                &      & 15/2 & 19276.94 & 19295 & 1.32 \\
$4f^{10}5d6s6p$ & Odd  & 15/2 & 24112.04 & 23908 & 1.35 \\
                &      &      &          &       &      \\
$4f^{10}5d6s^2$ & Even & 15/2 &  8427.11 &  8395 & 1.28 \\
                &      & 13/2 &  9147.08 &  9341 & 1.33 \\
$4f^{10}5d6s^2$ & Even & 15/2 & 12339.04 & 12903 & 1.23 \\
                &      & 13/2 & 12344.55 & 12953 & 1.23 \\
                &      & 11/2 & 13082.93 & 13799 & 1.25 \\
                &      & 13/2 & 15081.12 & 16459 & 1.17 \\
                &      & 11/2 & 16937.43 & 16817 & 1.13 \\
$4f^{11}6s6p$   & Even & 15/2 & 15855.28 & 15913 & 1.28 \\
                &      & 13/2 & 17059.35 & 17135 & 1.20 \\
$4f^{10}5d^26s$ & Even & 15/2 & 20167.17 & 20138 & 1.41 \\
\end{tabular}
\footnotetext[1]{NIST, Ref.~\cite{NIST}}
\end{ruledtabular}
\end{table}

Tables \ref{Dy-en} and \ref{Ho-en} present comparison between
experimental and theoretical energies and $g$-factors for Dy~I
and Ho~I atoms. The $g$-factors are useful for the identification
of the states and for control of configuration mixing~\cite{Dzuba02}.
For dysprosium atom both the energies and $g$-factors are reproduced
quite accurately. This includes the states with the energies of
19797.96~cm$^{-1}$ which are used in the
measurements~\cite{Budker07,Budker07a}.

For holmium the $g$-factors are not known for the most of the states.
This makes it more difficult to identify the states and to judge
about the accuracy of the calculations of the relativistic energy
shifts. If the measurements for holmium are to go ahead it would
be good to measure the $g$-factors as well, at least for the states
of most interest. At the moment we can only rely on the energies.
Although the energies are reproduced in the calculations quite
accurately the coefficients $q$ in (\ref{omega}) are very sensitive
to the configuration mixing which in turn is sensitive to the
energy intervals between close levels of the same parity and total
angular momentum. Therefore, having good accuracy for absolute values
of energies is not enough for reliable results for 
the coefficients $q$. It is very important that the relative
positions of the states around the states of interest are
reproduced accurately in the calculations.

\section{Results and discussion}

\begin{table}
\caption{Approximate values of the $q$-coefficients for different
configurations of Ho~I ($\times 10^3$ cm$^{-1}$).}
\label{Ho-q}
\begin{ruledtabular}
  \begin{tabular}{l l c r}
Configuration   & Parity & Transition\footnotemark[1] & \multicolumn{1}{c}{$q$} \\
\hline
$4f^{11}6s^2$   & Odd  &  (ground state)     &  5(5)~ \\
$4f^{10}5d6s^2$ & Even & $ 4f \rightarrow 5d$ & -35(15) \\
$4f^{11}6s6p$   & Even & $ 6s \rightarrow 6p$ &   4(4)~ \\
$4f^{10}6s^26p$ & Odd  & $ 4f \rightarrow 6p$ & -45(15) \\
$4f^{11}5d6s$   & Odd  & $ 6s \rightarrow 5d$ &  7(4)~ \\
\end{tabular}
\footnotetext[1]{single-electron transition from the ground state.}
\end{ruledtabular}
\end{table}

\paragraph{Holmium}
Holmium atom has been suggested for the search of the variation of
the fine structure constant by Mark Saffman~\cite{Saffman}.
From the computational point of view it represents the
most difficult case.
It has thirteen electrons in open shells, very dense spectrum,
strong configuration mixing and multiple level crossing
in the vicinity of the physical value of $\alpha$ when energies
are considered as functions of $\alpha^2$. All these factors contribute
to the instability of the results. Therefore, it is instructive
to start from simple estimations based on a single-electron
approximation. Table~\ref{Ho-q} shows approximate values of
$q$-coefficients for different configurations of holmium
obtained with the use of formula (\ref{qab}) but with
energy shifts of the individual single-electron states ($\Delta_{a,b}$)
taken from the Hartree-Fock calculations rather than from
formula (\ref{delta}). Note that we present the energies
and $q$-coefficients relative to the ground state. Therefore,
relativistic energy shift ($q$) for the ground state is zero
by definition. The $q$-coefficients for other states of the
$4f^{11}6s^2$ configuration are determined by the fine structure
of the $4f$ orbital.
Large error bars are due to the fact that
relativistic energy shifts depend on the values of the total
momentum $j$ of the single-electron states involved in the
transition (see formula (\ref{delta})). For example,
the $6s \rightarrow 6p$ transition can be the $6s \rightarrow 6p_{1/2}$
or the $6s \rightarrow 6p_{3/2}$ transition, etc.

As can be seen from the data in Table~\ref{Ho-q} the values of $q$
are very different for different configurations. For example,
$q \approx -35000 \rm{cm}^{-1}$ for the $4f^{10}5d6s^2$
configuration and $q \approx 4000 \rm{cm}^{-1}$ for the
$4f^{11}6s6p$ configuration. But these two configurations
have the same parity and can have the states of the same
total angular momentum $J$. Therefore, if these configurations are
strongly mixed the resulting values of $q$ will be  linear combinations of
$q \approx -35000 \rm{cm}^{-1}$ and $q \approx 4000 \rm{cm}^{-1}$, i.e.
they  may take any value between large negative value
and some positive one depending on which configuration dominates
in the state. The same is true for odd states which constitute
a mixture of the negatively shifted $4f^{10}6s^26p$ configuration
with the positively shifted $4f^{11}5d6s$ or $4f^{11}6s^2$
configurations. The analysis of the holmium spectrum shows that
there are many states of the same parity and total angular momentum $J$
which are separated by only small energy intervals and in which
different configurations dominate. These states are strongly mixed
which leads to instability of the calculations of the $q$-coefficients.
The only way to obtain reliable results is to make sure
that the relative position of the states in the vicinity of state
of interest as well as the energy intervals between these states
are reproduced accurately in the calculations. This can be achieved
by appropriate choice of the $\alpha_p$ parameters for different
configurations (see Table~\ref{sets}).

\begin{table*}
\caption{Experimental and theoretical energies and calculated
relativistic energy shifts ($q$-coefficients, cm$^{-1}$) for
some transitions of Ag~I, Dy~I, Ho~I, Yb~II, Yb~III, Au~I and Hg~II.}
\label{q-s}
\begin{ruledtabular}
  \begin{tabular}{l l l l l l r r r}
Atom &  \multicolumn{2}{c}{Ground state} &  \multicolumn{2}{c}{Excited state} &
\multicolumn{2}{c}{Energy[cm$^{-1}$]} &
\multicolumn{2}{c}{$q$-coefficients, [cm$^{-1}$]} \\
     & Conf. & $J$ & Conf. & $J$ & Expt.\footnotemark[1] & this work &
\multicolumn{1}{c}{this work} & \multicolumn{1}{c}{other} \\
\hline
Ag~I   & $4d^{10}5s$    & 1/2 & $4d^{9}5s^2$  & 5/2 & 30242.26~ & 30188 & -11300 &                     \\
       &                &     & $4d^{9}5s^2$  & 3/2 & 34714.16~ & 35114 &  -6500 &                      \\
Dy~I   & $4f^{10}6s^2$  &  8 & $4f^{10}5d6s$  & 10 & 19797.96~ & 20077 &   7952 &   6008\footnotemark[2] \\
       &                &    & $4f^{9}5d^26s$ & 10 & 19797.96~ & 19693 & -25216 & -23708\footnotemark[2] \\

Ho~I   & $4f^{11}6s^2$  & 15/2 & $4f^{10}5d6s^2$ & 11/2 & 20493.40 & 21763 &  -28200 & \\
       &                &      & $4f^{11}5d6s$   & 13/2 & 20493.77 & 20872 &    7300 & \\
       &                &      & $4f^{11}6s6p$   & 13/2 & 22157.86 & 22599 &    3000 & \\
       &                &      & $4f^{11}5d6s$   &  9/2 & 22157.88 & 22631 &    8000 & \\

Yb~II  & $4f^{14}6s$    & 1/2 & $4f^{13}6s^2$ & 7/2 & 21418.75~ & 20060 & -63752 & -56737\footnotemark[2] \\
       &                &     & $4f^{13}6s^2$ & 5/2 & 31568.08~ & 31303 & -53400 &                        \\
       &                &     & $4f^{13}5d6s$ & 5/2 & 26759.02~ & 26781 & -46863 &                        \\
       &                &     & $4f^{13}6s6p$ & 7/2 & 47921.31~ & 47927 & -60432 &                        \\
Yb~III & $4f^{14}$      &  0  & $4f^{13}5d$   &  0  & 45276.85~ & 46505 & -32800 & -27800\footnotemark[2] \\
Au~I   & $5d^{10}6s$    & 1/2 & $5d^{9}6s^2$  & 5/2 &  9161.3~~ &  9186 & -38550 &                        \\
       &                &     & $5d^{9}6s^2$  & 3/2 & 21435.3~~ & 22224 & -26760 &                        \\
Hg~II  & $5d^{10}6s$    & 1/2 & $5d^{9}6s^2$  & 5/2 & 35514.624 & 35066 & -52200 & -56670\footnotemark[3] \\
       &                &     & $5d^{9}6s^2$  & 3/2 & 50555.567 & 50886 & -37700 & -44000\footnotemark[3] \\
\end{tabular}
\footnotetext[1]{NIST, Ref.~\cite{NIST}}
\footnotetext[2]{Dzuba {\em et al}, Ref.~\cite{Dzuba03}}
\footnotetext[3]{Dzuba {\em et al}, Ref.~\cite{Dzuba99a}}
\end{ruledtabular}
\end{table*}

In Table~\ref{q-s} we present the results of the calculations for two
pairs of almost degenerate states of holmium. The relative change of frequency
between degenerate levels due to change of $\alpha$ is strongly
enhanced by small energy interval. The enhancement factor $K$ (defined by $\delta \omega/\omega=K \delta \alpha/ \alpha$
where $\omega$ is the transition frequency) is
given by~\cite{Dzuba99a}
\begin{equation}
  K = 2 \Delta q/\Delta E.
\label{enh}
\end{equation}
This enhancement factor is about $3 \times 10^5$ for both pairs
of holmium states presented in Table~\ref{q-s}. To avoid misunderstanding we should note
that the enhancement of the relative
effect here is due to the small $\Delta E$; there is no any enhancement of the absolute values of the frequency
shifts. The values of $q$ in holmium are typical for heavy atoms.

\paragraph{Dysprosium}
Dysprosium atom is used for the search of time-variation of the
fine structure constant at Berkeley~\cite{Budker07,Budker07a,Budker04}.
It has two almost degenerate levels of opposite parity at $E=19797.96 \rm{cm}^{-1}$
for which the enhancement factor (\ref{enh}) is about $10^8$~\cite{Dzuba03}.
Limits on the rate of changing of alpha in time obtained from
monitoring the frequency of the transition between these two levels
over long period of time is on the same level of precision as
for the most advances atomic optical clock experiments
($ \sim 10^{-15}/\rm{yr}$)~\cite{Budker07}. The interpretation of the
measurements are based on our early calculations~\cite{Dzuba03}.
The aim of present calculations is to check our previous result
with a significantly different method.

From the computational point of view dysprosium atom is an easier
case than holmium in two ways. First, it has one less valence
electron and its spectrum is much less dense. The energy separation
between mixing states is larger than 1000~cm$^{-1}$ which is much
easier to reproduce in the calculations than few hundred cm$^{-1}$
as in the case of holmium. Second, experimental values for $g$-factors
are available for dysprosium. The $g$-factors are almost as sensitive
to configuration mixing as the $q$-coefficients~\cite{Dzuba02}
providing an important test of the accuracy of the calculation.
As can be seen from Table~\ref{Dy-en} both energies and $g$-factors
are reproduced in the calculations with good accuracy.

The results for the $q$-coefficients are compared in Table~\ref{q-s}
with previous calculations. The largest relative difference is for
the smaller coefficient and is about 30\%. However, the difference
for $\Delta q$ which is  important for the interpretation of the
measurements is only 12\%.

\paragraph{Ag~I, Yb~II, Yb~III, Au~I and Hg~II}
The Ag~I, Yb~II, Yb~III and Hg~II atoms are also used or considered for the use
in the laboratory search for the variation of the fine structure constant (see~\cite{Lea}
and references therein). We have included Au~I because it has electron structure
similar to Ag~I and Hg~II. However, we are unaware about any plans to use Au in
the measurements.

All these systems utilize the use of a transition from the ground state
to a low lying state which involves an excitation from the core.
Both states have no significant admixture of other configurations,
relatively easy to calculate and produce stable results.

The results for the $q$-coefficients for the transitions of interest
are presented in Table~\ref{q-s}. Here we also have good agreement with
previous calculations for the cases when the data are available.

\section{Conclusion}

Calculations of the relativistic energy shifts are presented for
many transitions in many-valence-electrons systems which are used
or planned to be used in the laboratory search for variation
of the fine structure constant. Good agreement with previous
calculations confirms the analysis based on old results and
provides an estimate of the accuracy of the calculations.
Many atomic transitions are added which were never
considered before.

\section*{Acknowledgments}

We are grateful to D. Budker and M. Saffman for stimulating discussions.
The work was funded in part by the Australian Research Council.

\end{document}